\documentclass{article}
\usepackage[10pt]{extsizes}
\usepackage[T2A]{fontenc}
\usepackage[utf8]{inputenc}
\usepackage{cmap}
\usepackage{slashed}
\usepackage{hyperref}

\usepackage{amsmath}
\usepackage{amsfonts}
\usepackage{mathrsfs}
\usepackage{amssymb}

\allowdisplaybreaks[1]

\numberwithin{equation}{section}

\usepackage[svgnames]{xcolor}
\definecolor{refBlue}{rgb}{0.0, 0.0, 0.32}
\definecolor{urlBlue}{rgb}{0.0, 0.0, 0.46}
\usepackage{hyperref}
\hypersetup{
    colorlinks=true,
    linkcolor=refBlue,
    citecolor=refBlue,
    urlcolor=urlBlue,
    }

\begin{document}

\begin{titlepage}

\begin{center}
{\Large\bf Adjustment of Faddeev-Popov quantization to reducible gauge}\\
\vspace{0.1cm}
{\Large\bf
 theories: antisymmetric tensor fermion in $AdS_d$ space}

\vspace{1cm}
{\bf A.O. Barvinsky$^{1}$, I.L. Buchbinder$^{2,3,4}$, V.A. Krykhtin$^{5}$, D.V. Nesterov$^{1}$}

\vspace{0.5cm}
{\it $^1$I.E. Tamm Theory Department, P.N. Lebedev Physical Institute,\\
53 Leninsky Prospect, 119991, Moscow, Russia,\\
{\tt barvin@td.lpi.ru,\, nesterov@lpi.ru}\\
$^2$Bogoliubov Laboratory of Theoretical Physics,\\
 Joint Institute for Nuclear Research, 6, Joliot Curie, 141980 Dubna, Russia,\\
  {\tt buchbinder@theor.jinr,ru}\\
$^3$ Center of Theoretical Physics, Tomsk State Pedagogical University,\\
60 Kievskaya str., 634061, Tomsk, Russia, \\
$^4$ National Research Tomsk State University,
Lenin Av. 36, 634050, Tomsk, Russia,\\
$^5$ Tomsk Polytechnic University,
Lenin Av. 30, 634050, Tomsk, Russia\\
{\tt krykhin@tspu.ru}}

\vspace{0.5cm}

\end{center}

\begin{abstract}
We develop the method adjusting the Faddeev-Popov factorization procedure
for the quantization of generic reducible gauge theories with linearly dependent
generators and apply it to the first stage reducible model of second rank antisymmetric
fermion in $d$-dimensional AdS spacetime. The method consists in nested factorizations
of the gauge group volume for the determination of the consistently defined delta
function of reduced gauge conditions, group integration measure and gauge-fixed
contribution of ghosts. It is compared to the Batalin-Vilkovisky (BV) formalism of
quantizing theories with linearly dependent generators and shown to be equivalent
to it for first stage reducible theories. Nevertheless, the method under consideration,
unlike the BV formalism, from the very beginning leads to the functional integral with
fewer number of ghosts. Using this method we quantized the variant of fermionic totally antisymmetric
tensor-spinor theory in AdS space and derived its effective action in terms of the
functional determinants of special Dirac-type operators. Limitations of the method
are also discussed along with the prospects of its extension to higher reducibility
stages and higher rank models of antisymmetric fermions.
\end{abstract}

\end{titlepage}

\vspace{1mm}
\section{Introduction}
Gauge fields play a central role in modern theoretical high-energy
physics. Suffice it to say that the Yang-Mills gauge field is a
basic element of the Standard Model of fundamental interactions.
General relativity is also a gauge theory. By now, a large number of
different Fermi-Bose gauge theories have been constructed, the
structure of which differs significantly from the structure of the
Yang-Mills theory and general relativity. Therefore, developing
various gauge theories quantization methods is of paramount
importance.

The correct quantization of gauge theories of the Yang-Mills type
was first given in the famous works of Faddeev-Popov \cite{FP} and
DeWitt \cite{DeW} and has long ago been included in textbooks. These
works laid the foundation for powerful canonical BFV
(Batalin-Fradkin-Vilkovisky) \cite{FV}, \cite{BF} and covariant BV
(Batalin-Vilkovisky) \cite{Batalin:1981jr}, \cite{Batalin:1983ggl}
quantization methods for general gauge theories (for a detailed
description of these methods, see \cite{Henneaux:1992ig},
\cite{Gomis:1994he} and references therein).

A distinctive feature of general quantization methods is the
presence of a large number of different ghost fields, the exclusion
of which in concrete theories requires a certain amount of work. In
this regard, a simple Faddeev-Popov quantization method,
based on a rather intuitive disentanglement from the full functional
integral of the integration along the orbits of the gauge group,
has certain advantages since from the very beginning it leads to a
fewer number of ghost fields.
The justification of this method on the basis of canonical
quantization was later given by Faddeev \cite{F}. However, the use of
the Faddeev-Popov quantization method was limited only to
Yang-Mills type theories. To be more precise, the Faddeev-Popov
method yields a correct functional integral only for theories with
closed gauge algebra and independent gauge generators. When applied
to theories violating the second of these conditions, this method leads
to an ill-defined functional integral because the Faddeev-Popov
gauge fixing procedure fails and the corresponding
determinant becomes degenerate. Thus, it seems useful to adjust the
Faddeev-Popov method for the quantization of gauge theories with
linearly dependent generators (called reducible gauge theories).
In this paper we solve this problem for general gauge theories with a
closed algebra of gauge generators belonging to the first stage
of reducibility.

The first field theory with reducible generators was proposed
by Ogievetsky and Polubarinov (notoph theory) \cite{OP} and then re-opened in
\cite{KR} in the string context\footnote{Importance and significance
of the pioneering paper \cite{OP} is discussed in \cite{EIvanov}.}. Later,
quite a lot of various reducible gauge theories were constructed and
quantized by different methods. Note that many of these papers are related to
supersymmetric gauge theories formulated in superfield terms. The current
state of art with the relevant references in this area is presented
in detail in \cite{KuR}, to which we refer the reader.

The adjustment of Faddeev-Popov quantization to reducible gauge theories was already proposed
in \cite{BK} in context of N=1 superfield models coupled to supergravity background.
As an example, a quantization of notoph theory in curved space-time was also considered there.
Note that all the discussed models were Abelian. In the present paper we want to expound
this adjustment method for a much wider scope of models with first stage reducibility
and establish its links to the BV formalism accounting for the origin of BV extra-ghost,
ghosts for ghosts, etc., considered in much detail in \cite{BVred}.

As an application of the quantization method under consideration, we carry out
a quantization of the recently proposed model of the totally
antisymmetric rank $p$ tensor-spinors (fermionic $p$-forms) \cite{Buchbinder:2009pa},
\cite{Zinoviev:2009wh} \footnote{See also \cite{CFMS} where the
multi-symmetric reducible fermionic field models have been
studied.}. Such a model is formulated in AdS space and characterized by
reducible gauge transformations depending on $p$ and space-time
dimension and evidently can not be quantized by Faddeev-Popov method.
Quantization of this ferminic model has been recently
considered in flat spacetime \cite{Lekeu:2021oti} in the framework of BV formalism. In our opinion,
quantization of a free theory in a flat space, although interesting
in its own way, is not very indicative, since in this case the
effective action does not depend on any parameters and is simply a
constant. The aim of our paper is quantization in AdS space and derive the effective action for $p=2$ theory
in terms of
appropriate functional determinants.

The paper is organized as follows. In Sect.\ref{Sect2} we present this method which
is based on a sequence of nested Faddeev-Popov procedures of factorizing
the infinite volume of the invariance group in the definition of the
delta-function of reducible gauge conditions, group integration measure and the
construction of gauge-fixed ghost sector of the theory. In Sect.\ref{Sect3} we apply it to the
first stage reducible model of rank $p=2$ antisymmetric fermion in $d$-dimensional
AdS spacetime. Though we focus on the case $p=2$, in Conclusions we reflect
on a general case of arbitrary $p$ and discuss limitations of the method and
further prospects of this model.

\section{Faddeev-Popov quantization method for theories with linearly dependent generators\label{Sect2}}

\subsection{Conventional Faddeev-Popov method and its limitations}
Consider the theory with fields $\phi=\phi^i$ with the action $S[\phi]$. The action is invariant under the local gauge transformations $\phi\to\phi^f$, $S[\phi^f]=S[\phi]$. Here $(\phi^{i})^{f} = \phi^{i} +\delta\phi^{i}$, $\delta\phi^{i}= R^{i}{}_{\alpha}f^{\alpha}$, where $R^{i}{}_{\alpha}$ are the generators of gauge transformations and $f^{\alpha}$ are the gauge parameters\footnote{We use the DeWitt condensed notations where all continuous and discrete indices are included into the indices $i$ and $\alpha$.}. We assume that the gauge transformations form in general a non-Abelian closed algebra, gauge field $\phi$ and group parameters $f$ are bosonic, though the extension to the generic boson-fermion system is straightforward.

We begin with a brief description of the welll known Faddeev-Popov method of constructing the functional integral in the case when the gauge generators $R^{i}{}_{\alpha}$ are independent.  The main idea of the method consists in the insertion under the integration over $\phi$ the unity factor defined by the following integration over the group parameters
 \begin{eqnarray}\label{unity}
  \Delta_\chi[\phi]\int Df\,\delta[\,\chi(\phi^f))\,]=1.
 \end{eqnarray}
Here $\chi(\phi)$ denotes the set of gauge conditions gauging out the invariance of the action and $\Delta_\chi[\phi]$ is the Faddeev-Popov functional determinant in this gauge, which on the surface of enforced gauge conditions reads as
 \begin{eqnarray}
  \Delta_\chi[\phi]\,\Big|_{\,\chi(\phi)=0}={\rm Det}\,\Big(\frac{\delta\chi(\phi)}{\delta\phi}R(\phi)\Big).
 \end{eqnarray}

Then under the assumption of the group associativity, $(\phi^f)^h=\phi^{hf}$, and invariance of the functional measure of integration over the group, $Df=D(hf)$, (here $hf$ denotes the composition of group transfor\-mations with the parameters $f$ and then $h$) the Faddeev-Popov determinant turns out to be gauge invariant, $\Delta_\chi[\phi^f]=\Delta_\chi[\phi]$, which immediately allows one to disentangle from the naive ill defined functional integral an infinite volume of the group $\int Df=V=\infty$. This goes by changing the order of integrations and using the invariance of the integration measure in the space of $\phi$, $D(\phi^f)=D\phi$, along with the change of integration variable $\phi^f\to\phi$,
 \begin{eqnarray}\label{insertion}
  \int D\phi\,e^{iS[\phi]}&\to&\int D\phi\,e^{iS[\phi]}\Delta_\chi[\phi]\int Df\,\delta[\,\chi(\phi^f)\,]\nonumber\\
  &=&\int Df\, \int D\phi\,e^{iS[\phi]}\Delta_\chi[\phi]\,\delta[\,\chi(\phi)\,]
  =V\times \int D\phi\,e^{iS[\phi]}\Delta_\chi[\phi]\,\delta[\,\chi(\phi)\,].
 \end{eqnarray}
Here $V$ is a formal infinite group volume. The last factor in (\ref{insertion}) becomes a needed and well-defined Faddeev-Popov integral which is actually independent of the choice of gauge $\chi$. We emphasize once again that the above method of constructing the correct functional integral works well only if the gauge generators are independent.

When applied to a gauge theory with linearly dependent generators this method would break at several points. To begin with, this linear dependence implies that the gauge generators
$R^i_\alpha(\phi)$, considered as matrices with the indices $i$ of fields $\phi^i$ and gauge indices $\alpha$, that is the indices enumerating the gauge parameters $f^\alpha$, have their own right zero vectors $Z^\alpha_a$ enumerated by another set of indices $a$,
\begin{equation}\label{ReducibilityId}
R^i_\alpha Z^\alpha_{\,a}= 0,\qquad \alpha=1,...m_0,\qquad a= {1,...\, m_1 < m_0},
\end{equation}
where $m_0$ and $m_1$ denote respectively formal dimensionalities of the original gauge algebra and the space of $Z^\alpha_a$, the latter being linearly independent for stage one reducible gauge theories. This means that the actual number of local gauge symmetries is $m_0-m_1$ (the rank of the matrix $R^i_\alpha(\phi)$), rather than $m_0$. Therefore the actual number of independent gauge conditions to fix these symmetries should also be $m_0-m_1$. They can be represented by the redundant (or reducible) set of gauge conditions $\chi^\mu$ satisfying the condition of linear dependence with some left zero vectors $\bar Z^b_\mu$,
\begin{equation}\label{left_zero}
\bar Z_\mu^{b}\chi^\mu= 0,\qquad \mu=1,...m_0,\qquad b= {1,...\, m_1 < m_0}.
\end{equation}
Therefore, the conventional delta function of these linearly dependent gauge conditions, $\delta^{(m_0)}[\,\chi\,]=\prod_\mu \delta[\,\chi^\mu\,]\propto\delta^{(m_1)}(0)$, becomes ill defined.

The second problem is that the group integration measure $Df=\prod_\alpha Df^\alpha$ in (\ref{unity}) also becomes ill defined because the actual integration should run over $(m_0-m_1)$-dimensional group space rather than the $m_0$-dimensional one. Finally, the Faddeev-Popov determinant
 \begin{equation}\label{FP}
\Delta_\chi={\rm Det}\,Q^\mu_\alpha, \qquad Q^\mu_\alpha\equiv \frac{\delta\chi^\mu}{\delta\phi^i}R^i_\alpha,
\end{equation}
becomes vanishing in view of the right zero vectors $Z^\alpha_a$ of the Faddeev-Popov operator $Q^\mu_\alpha$, $Q^\mu_\alpha Z^\alpha_a=0$.

All these three difficulties can be successfully resolved by a simple method alternative to BV quantization scheme, which we develop in what follows. The idea of the method consists in the successive insertion of the unity factor, analogous to (\ref{unity}), and factorization of the group volume with a corrected measure $V=\int D\mu_f$.

\subsection{Delta function of reducible gauge conditions}
The first step is the construction of the correctly defined delta function of reducible gauge conditions. Naive delta function
\begin{equation}\label{bad-delta}
\delta[\chi]=\int D\pi\, e^{i\pi_\mu\chi^\mu}
\end{equation}
is ill defined because the integral over the Lagrange multiplier $\pi_\mu$ requires the factorization of the volume $V_1=\int D\xi_1$ of the invariance group of the exponential,
\begin{equation}\label{pi-invariance}
\pi_\mu\to\pi_\mu^{\xi_1}=\pi_\mu+\bar Z^b_\mu\xi_{1\,b}, \qquad\pi_\mu^{\xi_1}\chi^\mu=\pi_\mu\chi^\mu.
\end{equation}
This is the $m_1$-dimensional gauge invariance in the $m_0$-dimensional space of $\pi_\mu$, following from the reducibility (\ref{left_zero}) of gauge conditions $\chi^\mu$. Such a factorization proceeds by gauge fixing $\pi_\mu$ with the degenerate (delta-function type) gauge $\sigma_b(\pi)=\pi_\mu\sigma^\mu_b$ and inserting into (\ref{bad-delta}) the unity factor of the form analogous to (\ref{unity}),
\begin{eqnarray}\label{unity1}
  \bar\Delta_1\int D\xi_1\,\delta[\,\sigma(\pi^{\xi_1}))\,]=1,\qquad \bar\Delta_1\,\big|_{\,\sigma(\pi)=0}={\rm Det}\,\big(\bar Z_\mu^a\sigma^\mu_b\big).
 \end{eqnarray}
Similarly to (\ref{insertion}) this leads to
\begin{eqnarray}\label{insertion-gauge}
\int D\pi\, e^{i\pi_\mu\chi^\mu}&\to&\int D\pi\,e^{i\pi_\mu\chi^\mu}\bar\Delta_1\int D\xi_1\,\delta[\,\sigma(\pi^{\xi_1})\,]\nonumber\\
  &=&\int D\xi_1\, \int D\pi\,e^{i\pi_\mu\chi^\mu}{\rm Det}\,\bar Q\,\delta[\,\sigma(\pi)\,]
  =V_1\times {\rm Det}\,\bar Q\int Dc\,\delta[\,\chi^\mu+\sigma^\mu_b c^b\,],
\end{eqnarray}
where we expressed $\delta[\,\sigma(\pi)\,]$ as the integral over an auxiliary field $c^a$ and took the integral over $\pi_\mu$. This allows one to identify the correct expression for the delta function of reducible gauge conditions
\begin{eqnarray}\label{correct-delta}
\hat\delta[\,\chi\,]={\rm Det}\,\bar Q\int Dc\,\delta[\,\chi^\mu+\sigma^\mu_b c^b\,],\qquad \bar Q\equiv\bar Q^a_b=\bar Z_\mu^a\sigma^\mu_b.
\end{eqnarray}

Here we get the extra Faddeev-Popov operator $\bar Q$ involving the zero vector of reducible gauge conditions $\chi^\mu$. Note that the argument of the delta function $\delta[\,\chi^\mu+\sigma^\mu_b c^b\,]\equiv\prod_\mu \delta[\,\chi^\mu+\sigma^\mu_b c^b\,]$ is no longer reducible, and this delta function is well defined. Moreover, the gauge $\sigma_a(\pi)$ for $\pi_\mu$, that is the matrix $\sigma^\mu_b$ above, can be freely changed without altering the ``correct'' delta function. This can be checked by direct variation of (\ref{correct-delta}),
\begin{eqnarray}\label{}
\delta_\sigma\big(\hat\delta[\,\chi\,]\big)={\rm Det}\,\bar Q\int Dc\,\delta\sigma^\mu_b\Big[\,c^b\frac\delta{\delta z^\mu}+\bar Q^{-1b}_{\;\;\;\;c}\bar Z^c_\mu\,\Big]\,\delta[\,z\,]\,\Big|_{\,z^\nu=\chi^\nu+\sigma^\nu_a c^a}=0.
\end{eqnarray}
Here we took into account that for an invertible operator $\bar Q^a_b$ (whose invertibility we certainly assume for an admissible choice of $\sigma^\mu_b$) the following relation holds $c^b=\bar Q^{-1b}_{\;\;\;\;c}\bar Z^c_\alpha z^\alpha$
at $z^\nu=\chi^\nu+\sigma^\nu_a c^a$ in view of $\bar Z^c_\alpha\chi^\alpha=0$, and $z^\alpha(\delta/\delta z^\mu)\delta[z]=-\delta^\alpha_\mu\delta[z]$.

\subsection{Group integration measure}
The second step is the construction of group integration measure $D\mu_f$ by the factorization of the infinite volume of the $f_1^a$-integration. This type of divergent integration arises when one integrates over $f$ any functional of the transformed gauge field $\varPhi[\,\phi^f]$, because as a function of $f$ it is constant on the  orbit of $f_1$-transformation in the space of $f^\alpha$,
\begin{equation}\label{f1-symmetry}
f\to f^{f_1}, \quad f^\alpha\to f^\alpha+Z^\alpha_a f^a_1, \quad \phi^{f^{f_1}}=\phi^f.
\end{equation}
In other words, gauge parameters themselves become gauge fields subject to $f_1$-transformations with the new generators $Z^\alpha_a$ -- zero vectors of the original generators $R^i_\alpha$, see (\ref{ReducibilityId}). This factorization is achieved by using a new set of gauge conditions $\omega^a(f)=\omega^a_\alpha f^\alpha$ on $f^\alpha$. The insertion of the new unity
\begin{eqnarray}\label{unity1}
  \Delta_1\int Df_1\,\delta[\,\omega(f^{f_1}))\,]=1,\qquad \Delta_1\,\big|_{\,\omega(f)=0}={\rm Det}\,\big(\omega^a_\alpha Z^\alpha_b\big),
\end{eqnarray}
into the group integral gives by the same pattern
 \begin{eqnarray}\label{insertion-group}
  \int Df\,\varPhi[\,\phi^f]&\to&\int Df\,\varPhi[\,\phi^f]\,\Delta_1\int Df_1\,\delta[\,\omega(f^{f_1}))\,]\nonumber\\
  &=&\int Df_1\, \int Df\,\Delta_1\,\delta[\,\omega(f)\,]\,\varPhi[\,\phi^f]
  =V_1\times \int Df\,\Delta_1\delta[\,\omega(f)\,]\,\varPhi[\,\phi^f].
 \end{eqnarray}
Therefore, the group integration measure can be identified with
\begin{eqnarray}\label{correct-group}
D\mu_f=Df\,{\rm Det}\, Q_1\,\delta[\,\omega(f)\,],\qquad Q_1\equiv Q^{\;\,a}_{1b}=\omega^a_\alpha Z^\alpha_b.
\end{eqnarray}
Here the new Faddeev-Popov operator $Q_1$ is defined in terms of zero vectors of the original gauge generators, and it is also assumed to be invertible by the choice of gauge functions $\omega^a_\alpha$ in (\ref{unity1}).

\subsection{Functional integral for first stage reducible gauge theories}
Now we are ready to implement the procedure (\ref{insertion}) for first stage reducible gauge theories. First we determine the overall Faddeev-Popov operator $\Delta$ from the definition of unity with the corrected delta function of the gauge (\ref{correct-delta}) and the corrected group integration measure (\ref{correct-group}). The equation for $\Delta$ then reads
\begin{eqnarray}\label{unity2}
1=\Delta \int D\mu_f\,\hat\delta[\,\chi(\phi^f)\,]=\Delta\,{\rm Det}\,\bar Q\,{\rm Det}\,Q_1
\int Df\,Dc\,\delta[\,\omega^a_\alpha f^\alpha\,]\,\delta[\,\chi^\mu(\phi)+Q^\mu_\alpha f^\alpha+\sigma^\mu_b c^b],
\end{eqnarray}
where $Q^\mu_\alpha$ is a naive (degenerate) Faddeev-Popov operator having right and left zero vectors
\begin{eqnarray}\label{Q}
  Q^\mu_\alpha=\frac{\delta\chi^\mu}{\delta\phi^i}R^i_\alpha,\quad Q^\mu_\alpha Z^\alpha_a=0,\quad \bar Z^b_\mu Q^\mu_\alpha=0.
\end{eqnarray}
The integral of the product of two delta functions in (\ref{unity2}) equals the inverse of the following Jacobian
\begin{eqnarray}\label{Jacobian}
  \frac{D\big(\chi+Q f+\sigma c,\,\omega f\big)}{D\big(f,\,c\big)}=
  {\rm Det}\begin{bmatrix}
            \;Q^\mu_\alpha & \sigma^\mu_b \\
\\ \omega^a_\alpha & 0\;
        \end{bmatrix}={\rm Det}\,F^\mu_\alpha,
\end{eqnarray}
where $F^\mu_\alpha$ is the non-degenerate operator obtained from $Q^\mu_\alpha$ by adding the gauge-breaking term composed of the gauge matrices $\sigma^\mu_b$ and $\omega^a_\alpha$
\begin{eqnarray}\label{gfFP}
  F^\mu_\alpha=Q^\mu_\alpha+\sigma^\mu_a\omega^a_\alpha.
\end{eqnarray}

The last equality in (\ref{Jacobian}) easily follows from the identities for this operator $F=Q+\sigma\omega$ which we write in the index-free form. Multiplying $F$ from the right and from the left by the right and left zero vectors of $Q$ we get
\begin{eqnarray}\label{Ward}
  \bar Z\,F=\bar{Q}\,\omega, \quad F\,Z=\sigma\,{Q}_1,
\end{eqnarray}
where $\bar{Q}=\bar Z\sigma$ and $Q_1=\omega Z$ are defined respectively in Eq.(\ref{correct-delta}) and Eq.(\ref{correct-group}). Therefore, $\bar Z=(\bar Z\sigma)\,\omega F^{-1}$, whence $\omega\,F^{-1}\sigma=I\equiv\delta^a_b$. Multiplying the matrix in Eq.(\ref{Jacobian}) by another triangular matrix whose determinant is one, we get
\begin{eqnarray}\label{}
  {\rm Det}\left(\begin{bmatrix}
\;Q & \sigma \\
\omega & 0\;
\end{bmatrix}
\begin{bmatrix}
\;I & F^{-1}\sigma \\
 \omega & 0\;
\end{bmatrix}\right)= {\rm Det}\begin{bmatrix}
\;F & 0 \\
 \omega & I\;
\end{bmatrix}={\rm Det}\,F,
\end{eqnarray}
which proves Eq.(\ref{Jacobian}).

Thus, equation (\ref{unity2}) for $\Delta$ takes the form $1=\Delta\,{\rm Det}\,\bar Q\,{\rm Det}\,Q_1({\rm Det} F)^{-1}$, so that the full Faddeev-Popov determinant equals
\begin{eqnarray}\label{Ward}
  \Delta=\frac{{\rm Det} F}{{\rm Det}\,\bar Q\,{\rm Det}\,Q_1}.
\end{eqnarray}
Insertion of the unity (\ref{unity2}) into the functional integral by the pattern of Faddeev-Popov procedure (\ref{insertion}) then reads
\begin{eqnarray}\label{insertion3}
  \int D\phi\,e^{iS[\phi]}&\to&\int D\phi\,e^{iS[\phi]}\Delta\int D\mu_f\,\hat\delta[\,\chi(\phi^f)\,]\nonumber\\
  &=&\int D\mu_f\, \int D\phi\,e^{iS[\phi]}\,\hat\delta[\,\chi(\phi)\,]\,\frac{{\rm Det} F}{{\rm Det}\,\bar Q\,{\rm Det}\,Q_1}\nonumber\\
  &=&V\times \int D\phi\,Dc\,e^{iS[\phi]}\delta[\,\chi^\mu(\phi)+\sigma^\mu_a c^a]\,
  \frac{{\rm Det} F}{{\rm Det}\,Q_1}.
 \end{eqnarray}

As a result, after a formal factorization the functional integral for stage one reducible gauge theories equals
 \begin{eqnarray}\label{Z}
  Z=\int D\phi\,Dc\;e^{iS[\phi]}\,\delta[\,\chi^\mu(\phi)+\sigma^\mu_a(\phi) c^a]\,
  \frac{{\rm Det} F[\phi]}{{\rm Det}\,Q_1[\phi]}
 \end{eqnarray}
and coincides with that derived from a fundamental Batalin-Vilkovisky formalism \cite{Batalin:1983ggl}, which remains valid for field-dependent $Z^{\alpha}_a(\phi),\sigma^\mu_a(\phi), \omega^a_\alpha(\phi)$. Thus it successively passes the main requirement for the functional integral in gauge theories -- its on-shell independence of the choice of gauge conditions ingredients, that is the matrices $\sigma^\mu_a(\phi), \omega^a_\alpha(\phi)$ \footnote{Strictly speaking, the factorization of the infinite group volume $V=\int D\mu_f$ here is possible only in the case when this volume is a $\phi$-independent quantity. Otherwise, the change of the order of integration here is impossible, and $V=V[\,\phi\,]$ remains under the sign of integration over $\phi$. This is the case of simple Abelian theories with $\phi$-independent zero vectors $Z^{\alpha}_a$ and gauge conditions matrices $\sigma^\mu_a$ and $\omega^a_\alpha$. It seems, it can be possible also in some non-Abelian theories after dimensional type regularization of the infinite volume. In any case, this issue should be specially investigated in each specific theory. As a justification for the possibility of factorization of the above type can be the coincidence of the result
(\ref{Z}) with the corresponding result within the framework of the BV formalism.}.

Note that the field $c^a$ here is just the extra ghost introduced in \cite{Batalin:1983ggl}, ${\rm Det}\,F$ is the gauge-fixed ghost determinant and ${\rm Det}\,Q_1$ is the contribution of ghosts for ghosts. Note also that in the transition from the second line of Eq.(\ref{insertion3}) to (\ref{Z}) the determinant ${\rm Det}\,\bar Q$, involving the left zero vector $\bar Z^a_\alpha$, has completely cancelled out -- this property is fully consistent with the BV formalism in which this vector does not participate in the algebra generated by the BV master equation \cite{Batalin:1983ggl}.

To compare the expression (\ref{Z}) with that obtained in \cite{BVred}, we should mention that in \cite{BVred} it was explicitly derived in the class of nondegenerate gauges. The transition to these gauges can be done by regularizing the delta function $\delta[\,\chi^\mu+\sigma^\mu_a c^a]$ in terms of the Gaussian exponential function with some invertible and tending to infinity gauge-fixing matrix $\varkappa_{\mu\nu}$. Then, integration over the extra ghost field $c^a$ gives
 \begin{eqnarray}\label{nondegenerate}
  \int Dc\,\delta[\,\chi^\mu+\sigma^\mu_a c^a]&=&\big({\rm Det}\,\varkappa_{\mu\nu}\big)^{1/2}\int Dc\,e^{\frac{i}2(\chi^\mu+\sigma^\mu_a c^a)\varkappa_{\mu\nu}(\chi^\nu+\sigma^\nu_b c^b)}\nonumber\\
  &=&\left(\frac{{\rm Det}\,\varkappa_{\mu\nu}}{{\rm Det}\,\varkappa_{ab}}\right)^{1/2}e^{\frac{i}2\chi^\mu\varPi_{\mu\nu}\chi^\nu},\quad \varkappa_{\mu\nu}\to\infty,
 \end{eqnarray}
where $\varPi_{\mu\nu}$ is the orthogonal projection operator, $\varPi_{\mu\nu}\sigma^\nu_a=0$, built in terms of this matrix and its projection on the subspace of $\sigma$-gauge conditions $\varkappa_{ab}\equiv\sigma^\mu_a \varkappa_{\mu\nu}\sigma^\nu_b$,
\begin{eqnarray}
\varPi_{\mu\nu}=\varkappa_{\mu\nu}-\varkappa_{\mu\alpha}\sigma^\alpha_a\varkappa^{ab}\sigma^\beta_b\varkappa_{\beta\nu},\quad
\varkappa^{ab}=\big(\varkappa_{ba}\big)^{-1}.
 \end{eqnarray}
Thus, along with the preexponential factor, extra-ghost integration adds to the classical action the gauge breaking term, $S\to S+\tfrac12\chi^\mu\varPi_{\mu\nu}\chi^\nu$, which was derived in \cite{BVred}. The projector nature of the gauge-fixing matrix $\varPi_{\mu\nu}$ in this term implies that gauging the symmetries out is actually enforced, as it should be, by the irreducible part of gauge conditions among their full set $\chi^\mu$. Note that this set can be even overdetermined, that is not necessarily satisfying the linear dependence condition (\ref{left_zero}), as it was assumed in our derivation, -- the redundant part of $\chi^\mu$ is anyway projected out by $\varPi_{\mu\nu}$.

\section{Quantization of rank 2 antisymmetric fermion field in $AdS_d$ spacetime\label{Sect3}}
Let's now move on to quantization. The model of the second rank antisymmetric fermion in $AdS_d$ spacetime is described by the field $\psi_{\mu\nu}$ and its Dirac conjugated $\bar\psi_{\mu\nu}$ field with the action \cite{Buchbinder:2009pa}
\begin{eqnarray}\label{S}
S[\psi_{\mu\nu},\bar\psi_{\mu\nu}]
&=&
i\int d^dx\,g^{1/2}\bar{\psi}_{\mu_1\mu_2}
\gamma^{\mu_1\mu_2\sigma\nu_1\nu_2}\,D_\sigma\,\psi_{\nu_1\nu_2},
\quad
D_\mu=
\nabla_\mu\pm\frac{i\,r^{\frac{1}{2}}}{2}\;\gamma_{\mu},
\end{eqnarray}
where $\nabla_\mu$ is the Levi-Civita covariant derivative acting on the fermionic tensor, $\gamma^{\mu_1\ldots\mu_p}$ for any $p$ is the totally antisymmetric product of gamma matrices and $r$ is the modulus of the $AdS_d$ cosmological constant. This action is invariant under gauge transformations with the fermionic vector parameters $\lambda_\mu$ and $\bar\lambda_\mu$
\begin{eqnarray}
\delta\psi_{\mu_1\mu_2}
=D_{\mu_1}\lambda_{\mu_2}-D_{\mu_2}\lambda_{\mu_1},\quad
\delta\bar\psi_{\mu_1\mu_2}
=\bar D_{\mu_1}\bar\lambda_{\mu_2}-\bar D_{\mu_2}\bar\lambda_{\mu_1},
\label{dgt2-1}
\end{eqnarray}
($\bar D_{\mu}\bar\lambda_{\nu}\equiv \overline{D_{\mu}\lambda_{\nu}}$), while these transformations themselves are invariant under the first stage transformations with the spinor parameters $\lambda$ and $\bar\lambda$,
\begin{eqnarray}
\delta\lambda_{\mu}=D_{\mu}\lambda,\quad \delta\bar\lambda_{\mu}=\bar D_{\mu}\bar\lambda\equiv\overline{D_{\mu}\lambda},
\label{dgt2-0}
\end{eqnarray}
in view of the relations $D_{[\nu_1}D_{\nu_2]}\lambda\equiv0$,  $\bar D_{[\nu_1}\bar D_{\nu_2]}\bar\lambda\equiv0$ which are valid in $AdS_d$ spacetime.

Thus, this is the first stage reducible gauge theory subject to quantization formalism of the previous section, and the application of the above method is rather straightforward. Basic difference from the above bosonic formalism is that the fermion field, $\phi^i\mapsto (\psi_{\mu\nu},\bar\psi_{\mu\nu})$, and its gauge transformation parameters, $f^\alpha\mapsto(\lambda_\mu, \bar\lambda_\mu)$, $f^a_1\mapsto(\lambda, \bar\lambda)$, are of Grassman anticommuting nature, so that relevant functional integrals are of Berezin type and the resulting determinants get raised to the inverse power reflecting their superdeterminant nature.

Thus, in accordance with the bosonic case we impose on $\psi_{\mu\nu}$ the number $d$ (per spacetime point) of spinor gauge conditions reducible to $(d-1)$-independent ones, which in their turn correspond to the parameters $\lambda_{\mu}$ reducible to a transverse vector (we count only tensor components, their spinor dimensionality $2^{[d/2]}$ being implicitly included). We chose this set of gauges $\chi^\mu\mapsto \big(\chi_\mu(\psi),\bar\chi_\mu(\bar\psi)\big)$ as
\begin{eqnarray}
\chi_\mu(\psi_{\alpha\beta})=\gamma^\nu\psi _{\mu\nu}+\frac{1}{d}\gamma_\mu\gamma^{\alpha\beta}\psi_{\alpha\beta},
\quad
\gamma^\mu\chi_\mu=0.
\label{chi}
\end{eqnarray}
The gauge for the Dirac conjugated fermion $\bar{\psi}_{\mu\nu}$ is fully analogous -- in what follows we will formulate everything explicitly
for $\psi_{\mu\nu}$, while for $\bar{\psi}_{\mu\nu}$ the analogous formalism will be implicitly assumed.

The gauge \eqref{chi} is reducible like (\ref{left_zero}) with local gamma matrix $\gamma^\mu$ playing the role of left zero vectors $\bar Z^a_\mu$, $\bar Z^a_\mu\mapsto\gamma^\mu$ (the condensed index $a$ ranging over continuous spacetime coordinates $x$ and discrete spinor labels). Since it has $(d-1)$ independent components, one could have inserted in the functional integral the $(d-1)$-dimensional delta function $\delta^{(d-1)}[\chi]=\prod_i\delta[\chi_i]$, $i=0,...d-1$, but this would comprise a noncovariant formalism. On the other hand, the use of the formal $d$-dimensional delta function in view of (\ref{chi}) would encounter the unregulated $\delta^{(d)}[\chi_\mu]\sim\delta(0)\;\delta^{(d-1)}[\chi_i]$. Similarly to (\ref{bad-delta})-(\ref{pi-invariance}) with $\pi_\mu\mapsto\bar\psi^\mu$, $\xi_{1b}\mapsto\bar\lambda\gamma_\mu$ this problem manifests itself as the divergence of the Fourier integral $\delta[\chi_\mu]=\int D\bar{\psi}^\mu \exp\{i\bar{\psi}^\mu\chi_\mu\}$ in view of its exponential being invariant under $\bar{\psi}^\mu\to\bar{\psi}^\mu_{(\lambda)}=\bar{\psi}^\mu+ \bar{\lambda}\gamma^\mu$ because of the gamma-traceless gauge (\ref{chi}).

Therefore, following the lines of Sect.2 we construct the covariant delta function of gauge conditions by using the ``gauge for gauge'' $\sigma_a(\pi)\mapsto \sigma(\bar\psi^\mu)=\bar\psi^\mu\gamma_\mu$, $\sigma^\mu_a\mapsto\gamma_\mu$, and applying the Faddeev-Popov factorization of the group volume. Repeating the steps of (\ref{insertion-gauge})
\begin{eqnarray}
\hat\delta[\chi_\mu]=
\int D\bar{\psi}^\mu \exp\{i\bar{\psi}^\mu\chi_\mu(\psi_{\alpha\beta})\}\;\delta[\bar{\psi}^\mu\gamma_\mu],\quad
\delta[\bar{\psi}^\mu\gamma_\mu]
=
\int D\psi \exp\{i\bar{\psi}^\mu\gamma_\mu\psi\},
\end{eqnarray}
 we get to (\ref{correct-delta}) with $c^a\mapsto\psi$
\begin{eqnarray}
\hat\delta[\chi_\mu(\psi_{\alpha\beta})]
=
\int D\psi \;\delta[\chi_\mu(\psi_{\alpha\beta})+\gamma_\mu\psi],
\label{dg-delta2}
\end{eqnarray}
where we omitted the Faddeev-Popov operator factor $({\rm Det}\,\bar Q)^{-1}$ because this operator is ultralocal in spacetime, $\bar Q^a_b=\bar Z^a_\mu\sigma^\mu_b\mapsto \gamma^\mu\gamma_\mu\delta(x,y)$ and ${\rm Det}\,\bar Q\sim\exp\big(\delta(0)(...)\big)$ can be discarded in dimensional regularization or absorbed in the irrelevant local measure.

Integration measure over the group transformation \eqref{dgt2-1}-\eqref{dgt2-0}, $D\mu_f\mapsto D\mu_{(\lambda)}D\mu_{(\bar\lambda)}$, follows from the factorization procedure (\ref{f1-symmetry})-(\ref{insertion-group}). It goes with $f^\mu\mapsto(\lambda_\mu,\bar\lambda_\mu)$ and $f_{1a}\mapsto(\lambda,\bar\lambda)$ by using the following choice of gauge condition functions $\omega^a(f)\mapsto (\gamma^\mu\lambda_\mu, \bar\lambda_\mu\gamma^\mu)$ which correspond to local gauge-fixing matrices $\omega^a_\alpha\mapsto\gamma_\alpha$ acting in both $\lambda$ and $\bar\lambda$ spinor sectors. Note that the indices $a$ enumerating zero vectors of gauge generators $Z^\mu_a f_1^a\mapsto(D_\mu\lambda,D_\mu\bar\lambda)$ are just the spinor indices in both of these sectors.  According to the general expression (\ref{correct-group}) the result reads
\begin{eqnarray}
D\mu_{(\lambda)}=D\lambda_\mu\,\Delta_0^{-1}\, \delta[\gamma^\mu\lambda_\mu],
\qquad D\mu_{(\bar\lambda)}= D\bar\lambda_\mu\,\Delta_0^{-1}\, \delta[\bar\lambda_\mu\gamma^\mu].
\end{eqnarray}
Here $\Delta_0^{-1}$ is the inverse of the Faddeev-Popov determinant (remember the Grassmann parity of the group parameters), corresponding in the terminology of \cite{BVred} to the contribution of ghosts for ghosts. The determinant ${\rm Det}\,Q_{1\,b}^{\;\;\;a}={\rm Det}\,\big(\omega^a_\alpha Z^\alpha_b\big)\mapsto\Delta_0$ in both ($\lambda,\bar\lambda)$-sectors equals
\begin{eqnarray}
\Delta_0\equiv{\rm Det}\,\Big[i\slashed{\nabla}\pm\frac{1}{2}\,r^{1/2}\,d\Big],
\end{eqnarray}
where the Dirac operator $\slashed{\nabla}=\gamma^\mu\nabla_\mu$ is acting on the tensor rank zero spinor.

In contrast to the procedure (\ref{unity2})-(\ref{gfFP}) of Sect.\ref{Sect2} leading to the gauge-fixed ghost operator we will calculate the functional integral of (\ref{unity2}) in the sector of $\psi_{\alpha\beta}$-field
\begin{eqnarray}
\Delta^{-1}
&=&
\int D\mu_{(\lambda)}\;\hat\delta\Big[\chi_\mu\big(\psi_{\alpha\beta}^{(\lambda)}\big)\Big]
=
\nonumber
\\
&=&
 \Delta_0^{-1}
\int D\bar{\psi}^\mu D\lambda_\mu \;\delta[\bar{\psi}^\mu\gamma_\mu]\;\delta[\gamma^\mu\lambda_\mu]
\exp\bigg\{i\psi^\mu\Big(\chi_\mu(\psi_{\alpha\beta})+\frac{\delta\chi_\mu(\psi^{(\lambda)}_{\alpha\beta})}{\delta\lambda_\nu}\lambda_\nu\Big)\bigg\}
\label{dDetFP}
\end{eqnarray}
by the transition to gamma-irreducible components of the integration variables
\begin{eqnarray}
&&\bar{\psi}^\mu=\bar{\phi}^\mu-\frac{1}{d}\,\bar{\phi}\,\gamma^\mu, \quad
\bar{\phi}^\mu\gamma_\mu=0, \quad
\bar{\psi}^\mu\gamma_\mu=\bar{\phi}
\\
&&\lambda_\mu=\xi_\mu-\frac{1}{d}\gamma_\mu\xi,\quad
\gamma^\mu\xi_\mu=0,\quad
\xi=\gamma^\mu\lambda_\mu
\end{eqnarray}
The integral \eqref{dDetFP} then takes the form
\begin{eqnarray}
\Delta^{-1}
&=&
\Delta_0^{-1}
\int D\bar{\phi}^\mu D\bar{\phi} D\xi_\mu D\xi\;
\delta[\bar{\phi}]\;\delta[\xi] \;
\times
\nonumber
\\
&&\qquad{}
\times
\exp\Bigl\{i\bar{\phi}^\mu\Bigl(\chi_\mu-i\slashed{\nabla}_1\xi_\mu+i\frac{d}{d-2}\nabla_\mu\xi\Bigr)
+i\bar{\phi}\xi\Bigr\},
\label{dDetFP-2+}
\end{eqnarray}
where the operator $i\slashed{\nabla}_p$ acting on the totally antisymmetric fermion of any rank $p$ is defined by the equation
\begin{eqnarray}\label{nabla-p}
i\slashed{\nabla}_p
\equiv
i\slashed{\nabla}\pm\frac{1}{2}\;r^{\frac{1}{2}}(d-2p).
\end{eqnarray}
By shifting the integration variable $\xi_\mu=\xi_\mu'-i(\slashed{\nabla}_1)^{-1}\chi_\mu$ and integrating over $\bar\phi$,  $\bar{\phi}^\mu$ and $\xi_\mu'$ one obtains
\begin{eqnarray}
\Delta^{-1}= \Delta_0^{-1} \Delta_1, \quad \Delta_1={\rm Det}\,\big(i\slashed{\nabla}_1\big).
\label{dDetFP-2++}
\end{eqnarray}
It is important that the functional determinant $\Delta_1$ should be calculated here on the functional space of $\gamma^\mu$-irreducible vector fermions $\varPsi_\mu$ satisfying the irreducibility constraint $\gamma^\mu\varPsi_\mu=0$.

The same result holds for the contribution of the $\bar\psi_{\alpha\beta}$-sector of the theory. Therefore, the determinant factor of the fermionic version of the expression (\ref{Z}) equals
 \begin{eqnarray}
  \frac{{\rm Det}\,Q_1}{{\rm Det} F}=\Delta^{-2}=\frac{\Delta_1^2}{\Delta_0^2}.
 \end{eqnarray}
The remaining functional integral of (\ref{Z})
\begin{eqnarray}
Z&=&\frac{\Delta_0^2}{\Delta_1^2}\int D\psi_{\mu\nu} D\bar{\psi}_{\mu\nu}D\psi D\bar{\psi}\nonumber
\\
&&\quad{}
\times
\delta[\chi_\mu(\psi_{\mu\nu})+\gamma_\mu\psi]
\;\delta[\bar{\chi}_\mu(\bar\psi_{\mu\nu})+\bar{\psi}\gamma_\mu]
\exp\Big\{iS[\psi_{\mu\nu},\bar\psi_{\mu\nu}]\Bigr\}
\end{eqnarray}
is easy to calculate in terms of the $\gamma^\mu$-irreducible components of all integration variables $\psi_{\mu\nu}$ and $\bar\psi_{\mu\nu}$. Decomposing them into these components $\Psi_{\mu\nu}$, $\Psi_\mu$ and $\Psi$ according to the equations
\begin{eqnarray}
\psi_{\mu\nu}=\Psi_{\mu\nu}
+\frac{1}{d-2}\Bigl(\gamma_\mu \Psi_\nu -\gamma_\nu \Psi_\mu\Bigr)
+\gamma_{\mu\nu}\,\Psi,\quad
\gamma^\mu\Psi_{\mu\nu}=0, \quad\gamma^\mu\Psi_\mu=0,
\end{eqnarray}
(similarly for $\bar\psi_{\mu\nu}$) one gets
\begin{eqnarray}
Z&=&
\frac{\Delta_0^2}{\Delta_1^2}\int D\Psi_{\mu\nu} D\bar{\Psi}_{\mu\nu}D\Psi_{\mu} D\bar{\Psi}_{\mu}D\Psi D\bar{\Psi}D\psi D\bar{\psi}\;
\times
\nonumber
\\
&&\qquad{}
\times
\delta[\Psi_\mu+\gamma_\mu\psi]
\;\delta[\bar{\Psi}_\mu+\bar{\psi}\gamma_\mu]
\exp\Big\{i\tilde S[\,\Psi_{\mu\nu},\Psi_{\mu},\Psi,\bar\Psi_{\mu\nu},\bar\Psi_{\mu},\bar\Psi\,]\Bigr\},
\label{dZ=Gamma}
\end{eqnarray}
where $\tilde S[\,\Psi_{\mu\nu},\Psi_{\mu},\Psi,\bar\Psi_{\mu\nu},\bar\Psi_{\mu},\bar\Psi\,]=S[\,\psi_{\mu\nu},\bar\psi_{\mu\nu}]$ is the reparametrization of the original action (\ref{S}). The integrand here is restricted to the surface of gauge conditions at which both $\psi$, irreducible $\Psi_\mu$ and their barred counterparts are vanishing, so that the integral takes the form
\begin{eqnarray}
Z=\frac{\Delta_0^2}{\Delta_1^2}
\int D\Psi_{\mu\nu} D\bar{\Psi}_{\mu\nu}D\Psi D\bar{\Psi}\;
\exp\Big\{i\tilde S[\Psi_{\mu\nu},\Psi_{\mu}=0,\Psi,\bar\Psi_{\mu\nu},\bar\Psi_{\mu}=0,\bar\Psi]\Bigr\}
\label{dZ=Gamma+}
\end{eqnarray}
where
\begin{eqnarray}
\tilde S[\Psi_{\mu\nu},\Psi_{\mu}=0,\Psi,\bar\Psi_{\mu\nu},\bar\Psi_{\mu}=0,\bar\Psi]
=
-2\bar{\Psi}^{\mu\nu} \big(i\slashed{\nabla}_2\big)\Psi_{\mu\nu}
-\frac{(d-1)!}{(d-5)!}\;\bar{\Psi} \big(i\slashed{\nabla}_0\big) \Psi,
\end{eqnarray}
and the operators $i\slashed{\nabla}_p$, acting on $\gamma$-ireducible antisymmetric fermionic $p$-forms, $p=0,2$, are defined by Eq.(\ref{nabla-p}). Taking the remaining Gaussian integral we finally get the one-loop contribution to the functional integral for second rank antisymmetric fermion fields on the $AdS_d$ background,
\begin{eqnarray}\label{Z2}
Z_{p=2}=\frac{\Delta_0^2}{\Delta_1^2}\times\Delta_2\Delta_0=
\frac{\Delta_2\;\Delta_{0}^3}{\Delta_{1}^2}.
\end{eqnarray}
This is a final result for effective action in the theory under consideration. The calculation of the functional determinants ${\Delta}_{2}, {\Delta}_{1}, {\Delta}_{0}$ on irreducible fermionic forms is a separate problem. Each of these functional determinants is related to the Dirac-type operator on AdS spacetime and can be exactly calculated, e.g., by the generic spectral technique on homogeneous spaces \cite{Com}.

\section{Conclusions}
Let us briefly summarize our results. We have constructed the adjustment of Faddeev-Popov quantization procedure to reducible gauge theories for the case of first stage reducibility. The main result is given by the relation (\ref{Z}). This result was applied to a quantization of the novel theory which is a model of the total antisymmetric rank 2 tensor-spinor field in AdS spacetime. This allowed us to derive the effective action of the model as a combination of special functional determinants (\ref{Z2}). They are the determinants of Dirac-type operators acting on gamma-irreducible fermionic $p$-forms in $AdS_{d}$ space, which are calculable by known spectral methods on homogeneous spaces.

From methodological viewpoint the suggested method of treatment for reducible gauge theories looks easier than the BV technique, because it is more transparent and does not appeal to BV-BRST machinery of anti-bracket formalism and master equation \cite{Batalin:1981jr,Batalin:1983ggl,Henneaux:1992ig,Gomis:1994he}. However, of course it is less fundamental and rigorous, just like the Faddeev-Popov factorization method, but its justification comes from the comparison with the BV formalism and the fact that the final result passes the test of its on-shell gauge independence \cite{BVred}.


We believe that the method formulated in this work for theories of first stage of reducibility can in principle be extended to higher stage theories, like $p>2$ antisymmetric fermions having the sequence of zero vectors, $Z^{a_n}_{b_{n-1}}$, $n=1,...p-1$, annihilating the vectors of the previous stage,
\begin{eqnarray}
Z^{a_m}_{b_{m-1}}Z^{b_{m-1}}_{b_{m-2}}=0.
\end{eqnarray}
However, it looks like that a level of complexities of such an extension might exceed those of BV formalism. In any case, extension of our approach to
higher stages of reducibility requires a separate consideration. For this reason, in a forthcoming paper \cite{BBKN} we are going to apply a regular BV technique for a quantization of antisymmetric fermion models with $p>2$ in order to confirm the obtained result (\ref{Z2}) for $p=2$ and to prove the educated guess that this result for arbitrary higher tensor ranks reads as
\begin{eqnarray}
Z_p=\frac{\Delta_p\;\Delta_{p-2}^3\;\Delta_{p-4}^5\cdots}
{\Delta_{p-1}^2\;\Delta_{p-3}^4\cdots},
\quad \Delta_p={\rm Det}\,\big(i\slashed{\nabla}_p\big),
\end{eqnarray}
where $\Delta_p$ are the determinants of the Dirac-type operators (\ref{nabla-p}) acting on the irreducible components of the fermionic $p$-forms.\footnote{We have already checked correctness of this result for $p$=3 case.}

\section*{Acknowledgements}
The research of A.O.Barvinsky and D.V.Nesterov 
was supported by the Russian Science Foundation grant No. \href{https://rscf.ru/en/project/23-12-00051/}{23-12-00051}. 


\end{document}